\documentclass[fleqn,usenatbib]{mnras}

\usepackage[T1]{fontenc}
\usepackage{ae,aecompl}

\usepackage{subcaption}
\usepackage{graphicx}	% Including figure files
\usepackage{amsmath}	% Advanced maths commands
\usepackage{amssymb}	% Extra maths symbols
\usepackage{color}
\usepackage[normalem]{ulem}

\definecolor{purple}{rgb}{0.5, 0, 0.5}

\newcommand{\tom}[1]{}

\newcommand{\comments}[1]{}%let's you comment things out without % symbol on every line.

\title[Impact of Microlensing on Standardising GLSNe Ia]{The Impact of Microlensing on the Standardisation of Strongly Lensed Type Ia Supernovae}

\author[M. Foxley-Marrable et al.]{Max Foxley-Marrable,$^{1}$\thanks{E-mail: max.foxley-marrable@port.ac.uk}
Thomas E. Collett,$^{1}$
Georgios Vernardos,$^{2,3}$
\newauthor{Daniel A. Goldstein,$^{4,5}$
David Bacon$^{1}$}\\
$^{1}$Institute of Cosmology and Gravitation, University of Portsmouth, Dennis Sciama Building, Burnaby Road, Portsmouth, PO1 3FX, UK\\
$^{2}$Kapteyn Astronomical Institute, University of Groningen, P.O. Box 800, 9700 AV Groningen, The Netherlands \\
$^{3}$Centre for Astrophysics \& Supercomputing, Swinburne University of Technology, PO Box 218, Hawthorn, Victoria, 3122, Australia\\
$^{4}$Department of Astronomy, University of California, Berkeley,
CA 94720, USA\\
$^{5}$Lawrence Berkeley National Laboratory, Berkeley, CA 94720, USA \\
}

\date{Accepted 2018 May 17. Received 2018 May 2; in original form 2018 February 21}

\pubyear{2018}

\begin{document}
\label{firstpage}
\pagerange{\pageref{firstpage}--\pageref{lastpage}}
\maketitle

\begin{abstract}
We investigate the effect of microlensing on the standardisation of strongly lensed Type Ia supernovae (GLSNe Ia). We present predictions for the amount of scatter induced by microlensing across a range of plausible strong lens macromodels. We find that lensed images in regions of low convergence, shear and stellar density are standardisable, where the microlensing scatter is {$\lesssim$ 0.15 magnitudes}, comparable to the intrinsic dispersion of for a typical SN Ia. These standardisable configurations correspond to {asymmetric lenses with an image located far outside the Einstein radius of the lens.} Symmetric and small Einstein radius lenses {($\lesssim 0.5$ arcsec)} are not standardisable. We apply our model to the recently discovered GLSN Ia iPTF16geu and find that the large discrepancy between the observed flux and the macromodel predictions from \citet{More:2016sys} cannot be explained by microlensing alone. Using the mock GLSNe Ia catalogue of \citet{GoldsteinTimeDel2017}, we predict that $\sim$ 22\% of GLSNe Ia discovered by LSST will be standardisable, with a median Einstein radius of 0.9 arcseconds and a median time-delay of 41 days. By breaking the mass-sheet degeneracy the full LSST GLSNe Ia sample will be able to detect systematics in $H_0$ at the 0.5\% level.
\end{abstract}


%%%%%%%%%%%%%%%%%%%%%%%%%%%%%%%%%%%%%%%%%%%%%%%%%%

%%%%%%%%%%%%%%%%% BODY OF PAPER %%%%%%%%%%%%%%%%%%
\begin{figure*}
\includegraphics[width=0.93\linewidth]{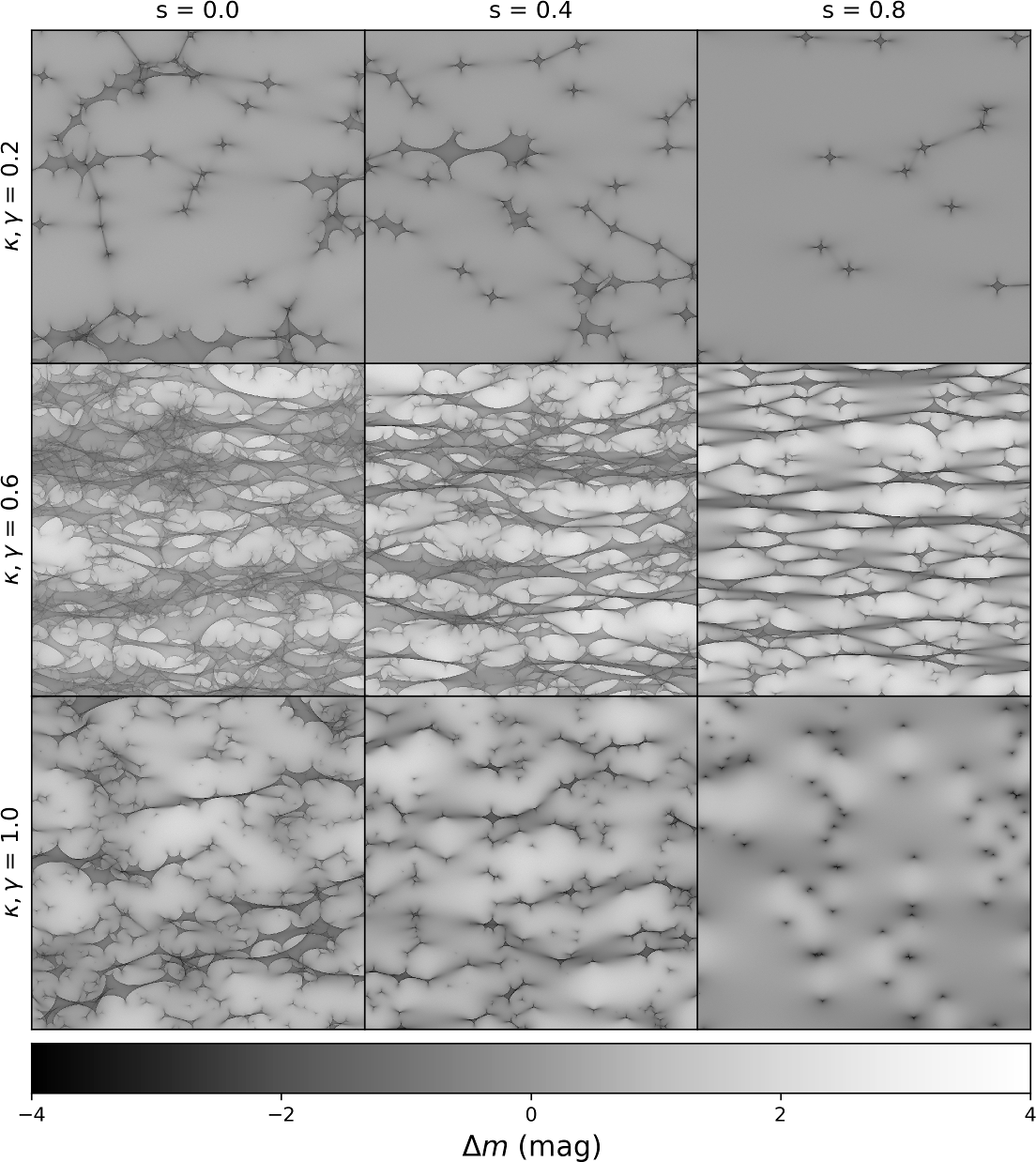}
\caption{{Example microlensing maps corresponding to various combinations of the convergence $\kappa$, shear $\gamma$ and smooth (dark) matter fraction $s$. Each side in a subpanel spans a physical range of $13.7R_{\mathrm{Ein},\odot}$} {($5 \times 10^{12}$ km)}. The maps show microlensing caustics projected onto the source plane as a result of inverse ray-tracing through a foreground star field. \comments{Background sources will be magnified/demagnified with amplitudes determined by the position and size of the light profile cross-section relative to the microlensing caustics.} The colour scale represents the deviation in magnitudes from the smooth macromodel magnification. \tom{Add scale bar}}
\label{GERLUMPHExample}
\end{figure*}

\begin{figure*}
\includegraphics[width=\linewidth]{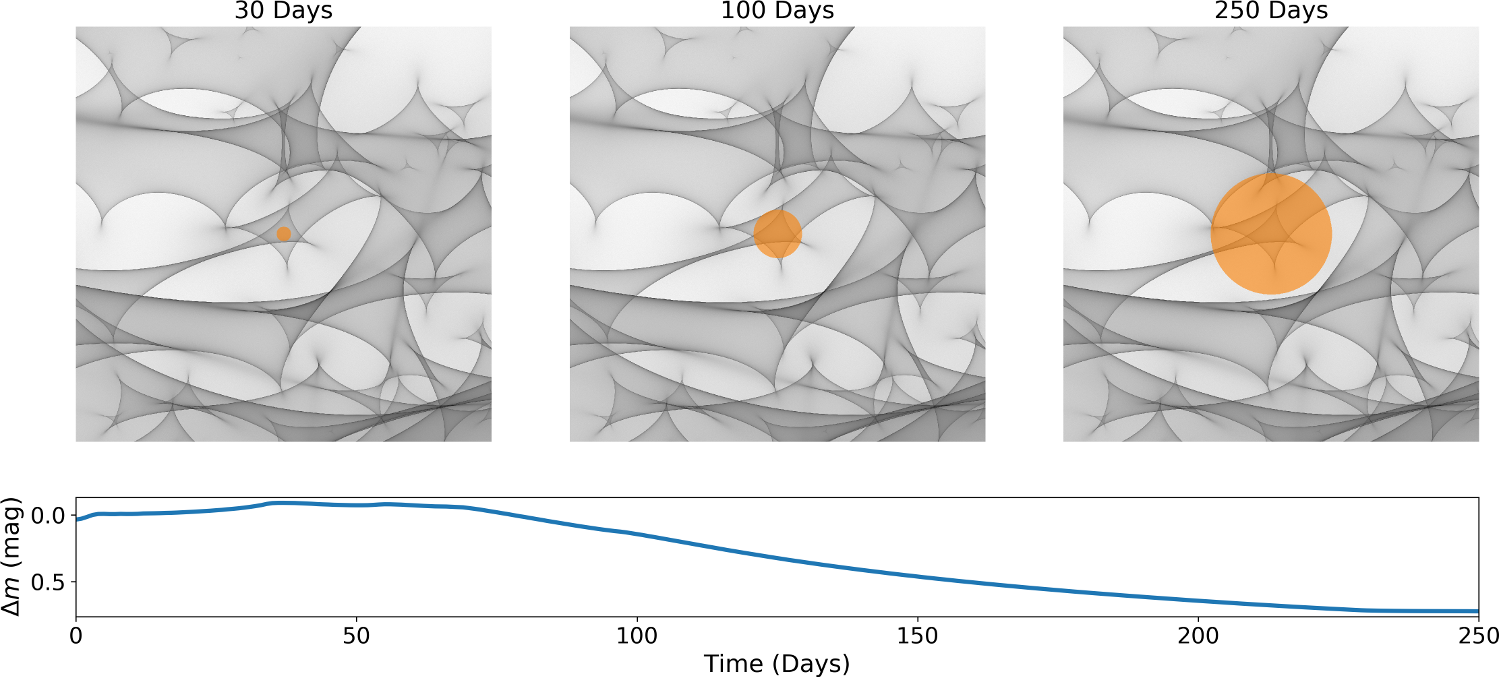}

\caption{A zoomed in microlensing map (each side corresponding to $2.75R_{\mathrm{Ein}, \odot} =$ {$1 \times 10^{12}$ km}) with a SN Ia profile superimposed on top, expanding at a rate of {$2.5 \times 10^{-8} R_{\mathrm{Ein}, \odot}$ s$^{-1}$ = $10^4$ km s$^{-1}$}. At each time step, the SN Ia profile is convolved with the microlensing background: the magnifications inside the disc are summed up and averaged. As the SN Ia profile grows, it crosses more and caustics, causing the microlensing magnification to vary over time. {The resulting magnifications are shown in the bottom panel.}}
\label{ConvolutionExample}
\end{figure*}

\section{Introduction} \label{sec:intro}

The value of the Hubble constant $H_0$ is a major point of contention in cosmology today, bringing the validity of the $\Lambda$CDM model of cosmology into question. This particularly arises from the $3.4\sigma$ tension between the \citet{Planck2016} result of $H_0 = 67.8 \pm 0.9$ km s$^{-1}$ Mpc$^{-1}$, derived from the cosmic microwave background (CMB), and the \citet{Riess2016} result of $H_0 = 73.2 \pm 1.7$ km s$^{-1}$ Mpc$^{-1}$, measured from low redshift supernovae and cepheids. The \citet{Planck2016} result infers $H_0$ assuming $\Lambda$CDM, whilst the \citet{Riess2016} result probes $H_0$ directly. Whilst this tension could be attributed to statistical fluke or unaccounted systematics, it potentially signals new physics beyond the $\Lambda$CDM model. Hence the need for precise and independent measurements of $H_0$ is greater than ever, such that the validity of the $\Lambda$CDM model can be tested.

Strong gravitational lenses are powerful probes of cosmology (\citealt{Oguri2012}; \citealt{Suyu2013}; \citealt{Collett2014}) and are particularly sensitive to $H_0$ through time delay cosmography \citep{TreuMarshall2016A&ARv..24...11T}. The light from each image in a lensing system takes a different path through the lens before reaching the observer. If the lensed object is a variable source, the images vary asynchronously with a geometrical time delay based on these path differences. Time delays have an additional component caused by the gravitational potential of the lens \citep{Shapiro1964PhRvL..13..789S}. When a background source peaks sharply in luminosity, the time delay between each image can in principle be measured by observing the time difference between the peaks of each image.

Time delays allow strong gravitational lenses to measure $H_0$ independently of assumptions made in the cosmological model: the value of $H_0$ is mostly invariant to other cosmological parameters such as the curvature and the dark energy equation of state \citep{H0LiCOWV2017}. The most recent example is \citet{H0LiCOWV2017}, which used time delays from three lensed quasars to independently measure $H_0 = 71.9^{+2.4}_{-3.0}$ km s$^{-1}$ Mpc$^{-1}$ to within a 3.8\% precision. In order to obtain a value for $H_0$ through strong lens time delays, one needs to know the 2D lens potential and the unlensed source position, neither of which can be observed directly. The use of lens modelling is therefore required in order to infer these quantities.

However strong lens models are subject to degeneracies, which are a major source of uncertainty for time-delay cosmography \citep{SchneiderSluse2014}. The main component of the degeneracy is the mass-sheet degeneracy: when rescaling the mass of the lens with an additional sheet of mass of constant density, the image configurations remain exactly the same but the projected mass on each image (also known as the convergence $\kappa$) changes, affecting the time delay \citep{Falco1985}. Put simply, two lens models producing identical image configurations can have very different time delays. Breaking the mass-sheet degeneracy is therefore necessary to constrain $H_0$. In order to break the mass-sheet degeneracy additional information is required, such as the intrinsic luminosity of the background source \citep{Kolatt1998MNRAS.296..763K}.

Originally proposed by \citet{Refsdal1964MNRAS.128..307R}, the prospect of using strongly lensed supernovae (GLSNe) to precisely measure $H_0$ is promising, especially after the discovery of the Type Ia GLSN iPTF16geu in October 2016 \citep{Goobar:2016uuf}. The light curves of Type Ia Supernovae (SNe Ia) are standardisable \citep{phillips1993apj}, allowing us to infer their intrinsic luminosity, hence GLSNe Ia can potentially lift the mass-sheet degeneracy \citep{Oguri2003MNRAS.338L..25O} and enable a test of systematic uncertainties in time delay cosmography.

{GLSNe are advantageous over the lensed active galactic nuclei (AGN) currently used for time delay cosmography (\citealt{Vuissoz2008A&A...488..481V}; \citealt{Suyu2010A&A}; \citealt{Tewes2013A&A...556A..22T}; \citealt{Bonvin2016A&A...585A..88B}). SN light curves have a strong peak before they decay, occurring over a time-scale of several weeks, whilst AGN light curves vary stochastically and heterogeneously, with weak variations in luminosity.} Hence GLSN time delays can be obtained in a single observing season, whilst AGN must be monitored over several years in order to acquire accurate time delays \citep{Liao2015ApJ...800...11L}.

\citet{GoldsteinTimeDel2017} predicts that $\sim$ 930 GLSNe Ia will be discovered by the Large Synoptic Survey Telescope (LSST) \citep{LSST2009arXiv0912.0201L} over its 10 year survey, with 70\% of the GLSNe Ia having time delays that can be measured precisely.

\begin{figure*}
\includegraphics[width=\linewidth]{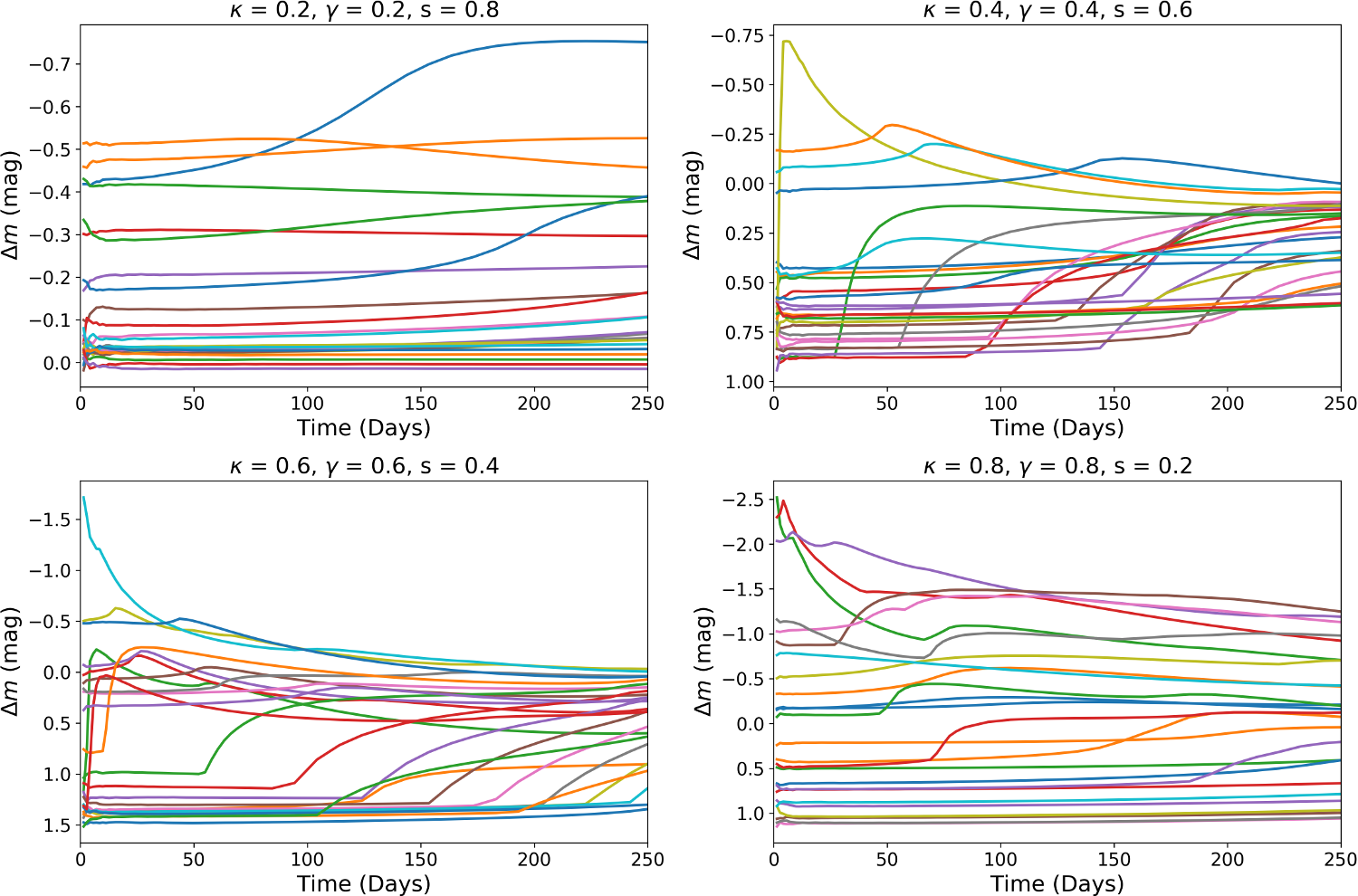}
\caption{A selection of simulated microlensing light curves for an expanding uniform disk. The four panels correspond to different values of the convergence $\kappa$, shear $\gamma$ and smooth matter fraction $s$.
Each light curve within a panel has the same macrolensing parameters but a different realisation of the microlensing by stars.}
\label{MicroLightCurves}
\end{figure*}

Despite the potential power of GLSNe Ia there exists one major theoretical barrier to their use as cosmological probes: microlensing caused by stars in the foreground lensing galaxy. Microlensing can independently magnify or demagnify individual images of the background source (\citealt{Dobler:2006wv}; \citealt{Bagherpour2006ApJ...638..946B}), introducing scatter into the shape and amplitude of the resulting light curves. The effect of microlensing on each lensed image can be inferred by obtaining its convergence $\kappa$\footnote{$\kappa$ is composed of both stellar and dark matter components.}, shear $\gamma$ and smooth matter fraction $s$ through lens modelling. $\kappa$ and $\gamma$ represents the amount of mass projected on and near the image respectively while $s$ represents the projected fraction of mass in dark matter as opposed to stellar matter (see Figure \ref{GERLUMPHExample}). Due to the distribution and random motion of the stars in the foreground galaxy, inferring the effect of microlensing on one image does not infer the effect of microlensing on the other image(s). This can significantly reduce the reliability of any time delay and luminosity measurement, as microlensing can randomly distort the light curve of each image, such that the intrinsic magnification and luminosity of the source can be difficult to determine. This effect also evolves over time. As the background SN Ia grows, the number of microlensing caustics that its light profile intersects with increases with time (see Figure \ref{ConvolutionExample}). Recently, \citet{GoldsteinNugentLSST2017} have shown that time delays can be robustly measured using early time colour curves.

\citet{More:2016sys} modelled iPTF16geu using the GLAFIC \citep{Oguri2010PASJ} and GLEE \citep{Suyu2010A&A, Suyu2012ApJ} macro lens models. Whilst the models themselves were in agreement, they were in contention with the \citet{Goobar:2016uuf} observations, with a discrepancy of almost 2 magnitudes for the brightest image. Their conclusion was that the disparity between their lens models and the observations was {primarily} due to microlensing from foreground stars in the lensing galaxy\footnote{{ \citet{More:2016sys} also mention the possibility of milli-lensing.}}.

In this article, we examine the effect microlensing has on an expanding SN Ia profile across a wide range of image configurations corresponding to particular values of $\kappa$, $\gamma$ and $s$. We provide the first predictions for regions of parameter space where the SN Ia image has a standardisable light curve, allowing us to infer its intrinsic luminosity and hence break the mass-sheet degeneracy. We define a standardisable supernova as one where the scatter due to microlensing is less than 0.15 magnitudes, comparable to the intrinsic dispersion for an unlensed SN Ia after standardisation (\citealt{Betoule2014}; \citealt{Macaulay2017}). We also present predictions for the fraction of GLSNe Ia discovered by LSST that will be standardisable. Finally we analyse the effect of microlensing on iPTF16geu and compare our results against the \citet{More:2016sys} prediction. In Section \ref{sec:sim} we describe our microlensing simulations. In Section \ref{sec:results} we present and discuss our subsequent analysis with results and conclude in Section \ref{sec:conclusion}. Throughout this paper we report results in the observer time frame assuming a source redshift of 0.409.

\begin{figure*}
\includegraphics[width=\textwidth]{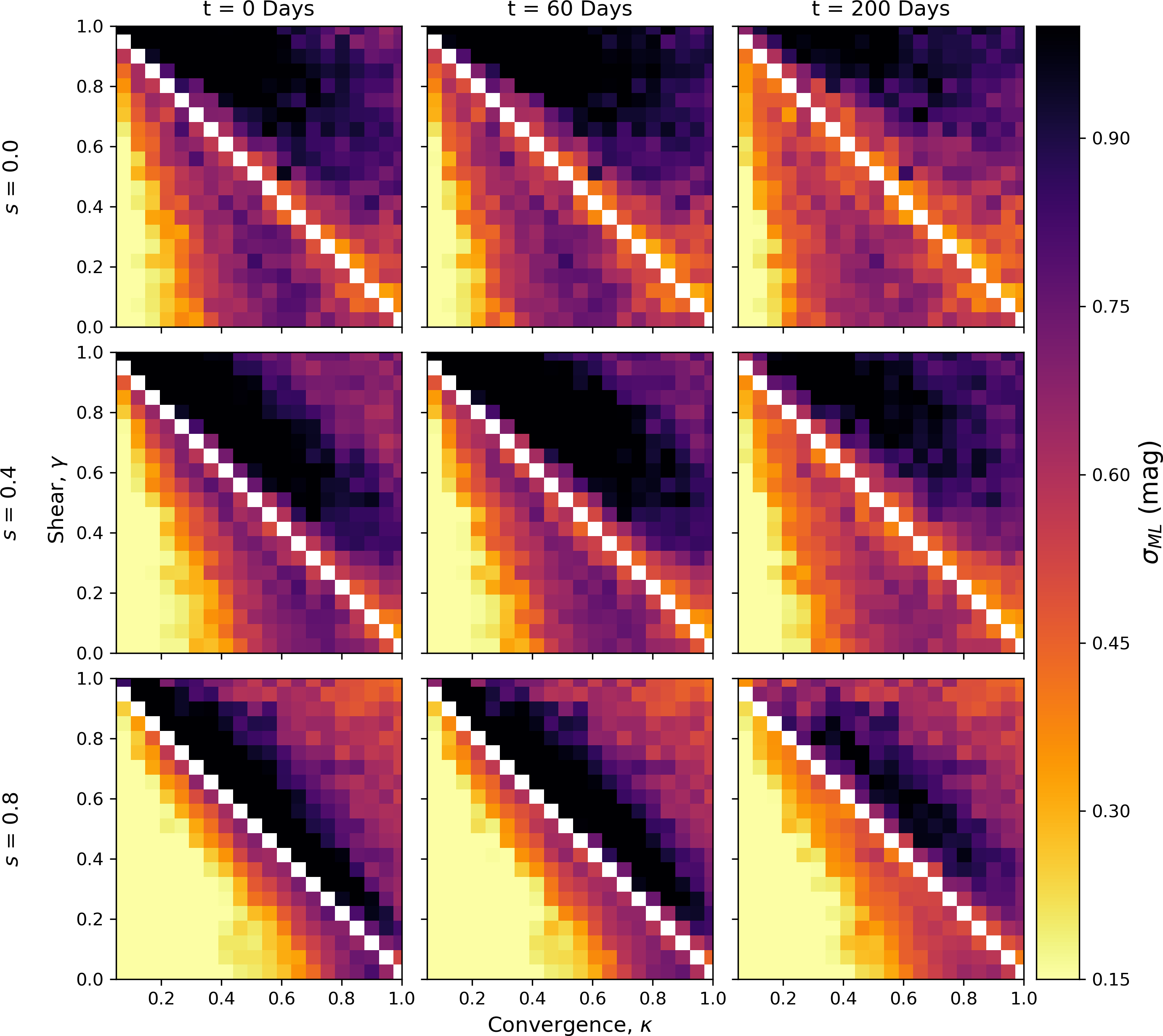}
\caption{The microlensing scatter in the observed luminosity of a lensed supernova image. Each subpanel shows the scatter as a function of the convergence $\kappa$ and shear $\gamma$ at fixed smooth matter fraction $s$ and time $t$ after explosion. The smooth matter fraction increases from top to bottom and the time of observation increases from left to right. {The white pixels along the diagonal correspond to regions of infinite magnification: lensed images do not form here.}}
\label{GD1scattermaps}
\end{figure*}

\begin{figure*}
\includegraphics[width=\linewidth]{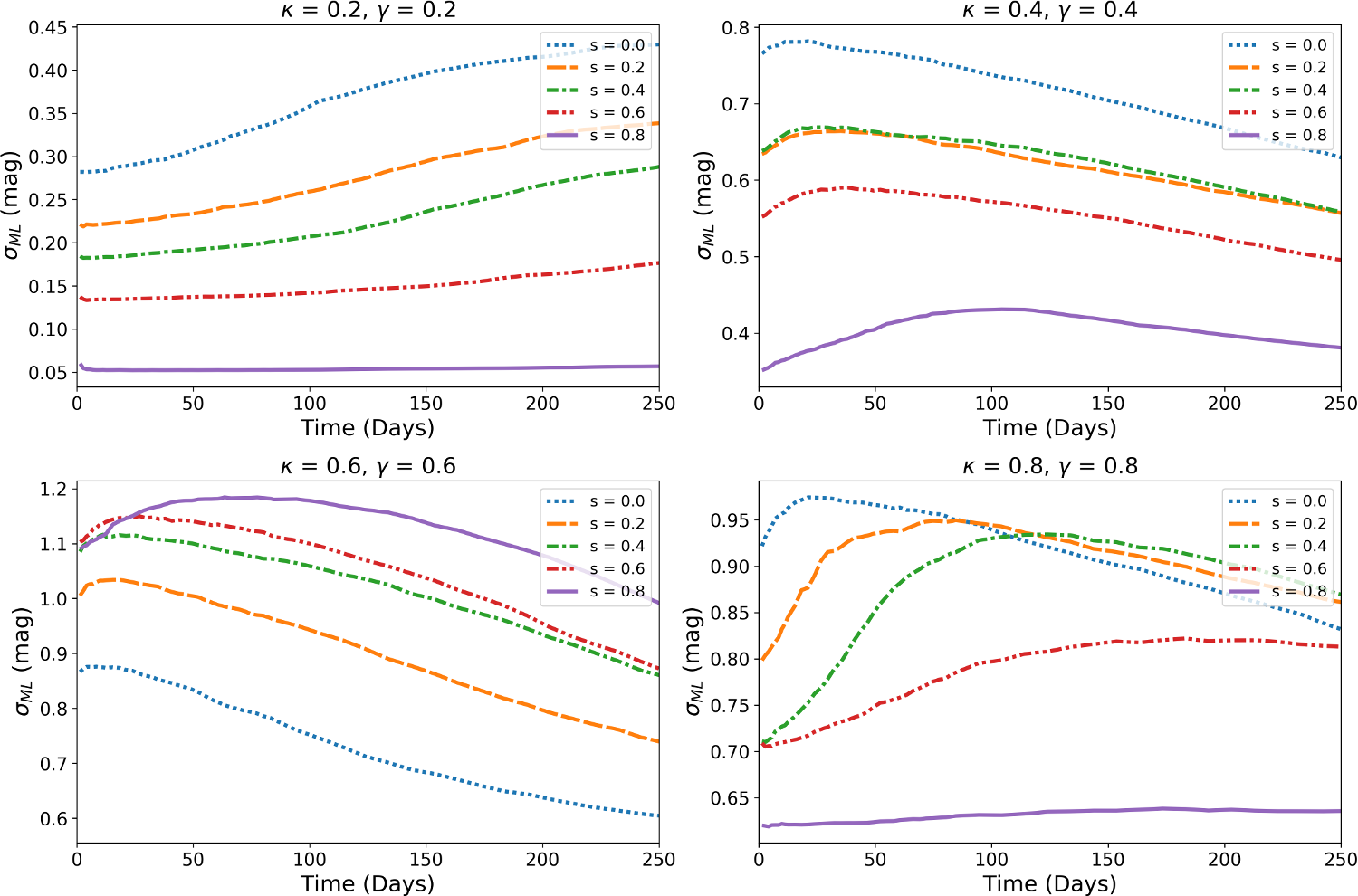}
\caption{The microlensing induced scatter in the observed luminosity as a function of time. The different lines correspond to different values of the smooth matter fraction $s$.}

\label{GD1Scatter}
\end{figure*}

\section{Simulations} \label{sec:sim}

To simulate the effect of microlensing by stars in the lens galaxy, we use magnification maps generated by the GERLUMPH project\footnote{\tt http://gerlumph.swin.edu.au} as shown in Figure \ref{GERLUMPHExample} \citep{Vernardos2014a,Vernardos2014b}.
These are pixelated maps of the source plane where the magnification per pixel has been calculated using the inverse ray-shooting technique \citep{Kayser1986}. A field of randomly distributed point masses is used to simulate the star field, with each star having the same mass.
The deflections for each microlens are computed directly and in parallel using the graphics processing unit (GPU) implementation of \citet{Thompson2010,Thompson2014}.

\comments{{\tom{I don't think we need this.} Integrate into section: The image separation is usually characterised in terms of the Einstein radius:
\begin{equation} 
R_{\mathrm{Ein}} = \sqrt{\frac{4GM}{c^2}\frac{D_{LS}}{D_L D_S}},
\label{eq:rein}
\end{equation}}
}

Each magnification map used in this work is square with 10000 pixels on a side, with a side corresponding to 13.7 $R_{\mathrm{Ein}, \odot}$. $R_{\mathrm{Ein}, \odot}$ is the Einstein radius for a 1 $M_{\odot}$ microlens; for a lens at $z_l=0.216$ and a source at $z_s=0.409$. This corresponds to a physical scale of {$R_{\mathrm{Ein}, \odot}$ = $4 \times 10^{11}$ km $= 2 \times 10^{-6}$ arcseconds on the sky, with each map pixel covering an area of $2.5 \times 10^{17}$ km$^2$ $= 9 \times 10^{-18}$ arcseconds$^2$ on the sky}.\footnote{Here we assume the best-fit $\Lambda$CDM cosmological parameters of \citet{Planck2016}.}

To sample a wide range of possible GLSN configurations we use the GD1 set of maps \citep[described in detail in][]{Vernardos2014a}. This set covers $\kappa, \gamma$ space on a uniform grid with $\Delta\kappa,\Delta\gamma = 0.05$ and $0 \lesssim \kappa,\gamma \lesssim 1.7$.
For each $\kappa,\gamma$ combination there are 11 values of $s$ available: $0 \leq s \leq 0.9$, in steps of 0.1, and $s=0.99$. For each $\kappa,\gamma$ pair we use maps with a smooth matter fraction $s=0.2,0.4,0.6,0.8$. This results in a total of 4488 magnification maps.

To obtain the SN microlensing light curves we convolve the magnification maps with a time varying profile of the background source (see Figure \ref{ConvolutionExample}). We use an expanding uniform disc to approximate the SN brightness profile. This simple model is sufficient for our purposes since the observed luminosity of a microlensed source is mostly sensitive to the average size and largely independent of any specific shape of the source profile \citep{Mortonson2005}. {We do not consider sources with clumpy profiles.}

The supernova expansion rate is set to $10^4$ km s$^{-1}$ {($2.5 \times 10^{-8} R_{\mathrm{Ein}, \odot}$ s$^{-1}$)}. Since this is much larger than any of the velocities involved (i.e. the peculiar velocity of the lens and the source, the velocity of the observer, and the proper motions of the microlenses), we can approximate the microlensing map as time invariant, and the centroid of a supernova as constant; only the radius of the supernova changes with time. At this expansion rate we are able to place $10^4$ SNe per magnification map without profile overlap within the first 60 days.

To obtain a light curve for an individual SN, we choose a position on the magnification map, and evaluate the product of its profile and the magnification map at each time step.
This is done for a total of 55 time steps: $0 < t < 16$ days with $\delta t = 1$ day, $16 < t < 60$ days with $\delta t = 2$ days, and $60 < t < 200$ days with $\delta t = 7$ days. 
A total of $\approx250,000$ convolutions between maps and profiles have been performed, requiring roughly 300 GPU hours.

We normalize all of our unlensed source fluxes to unity  at all times, such that our light curves depend only on the microlensing rather than intrinsic variations in the unlensed source. We show a range of example microlensing light curves in Figure \ref{MicroLightCurves}. 
%\footnote{Given in terms of $\Delta m$ i.e. the deviation in magnitudes from the smooth macromodel magnification $\mu_\mathrm{smooth}$ as a result of microlensing.}.}

\section{Results and Discussion} \label{sec:results}
\begin{figure*}
\includegraphics[width=\linewidth]{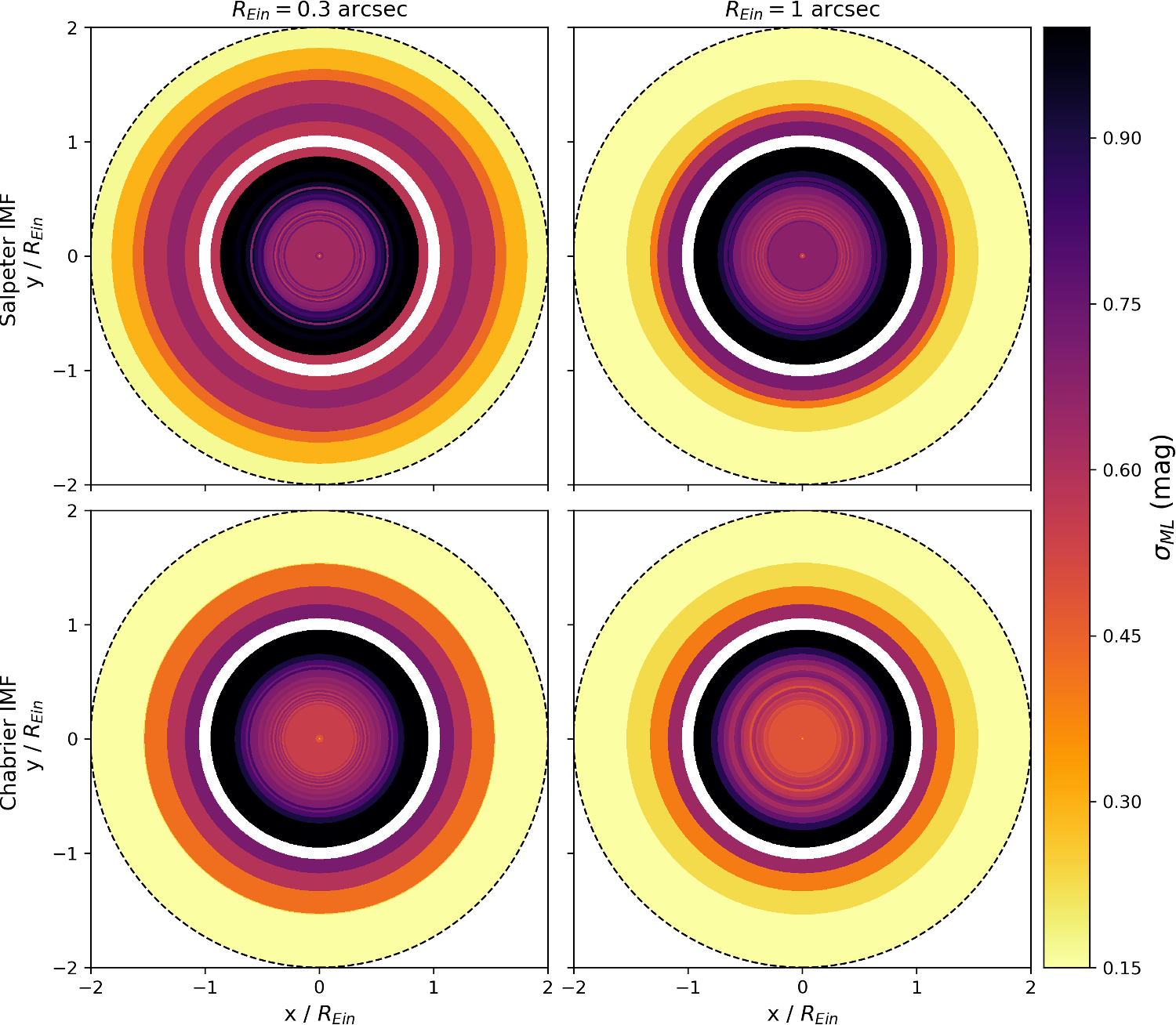}
\caption{{The amount of microlensing scatter induced on any lensed image at any point on the image plane, assuming an SIS lens and a SN 30 days after explosion. The four panels show the effect of varying the Einstein Radii $R_\mathrm{Ein}$ and the Initial Mass Functions. The black dashed line represents the outermost boundary for multiple imaging. {The inner white circle corresponds to the critical curve and infinite magnification.} The Salpeter and Chabrier IMFs correspond to the dark matter fractions derived in \citet{SLACS_X}.}}
\label{ImgScatter}
\end{figure*}

\subsection{Microlensing Scatter} \label{sec:Micro_Scatter}
For each $\kappa-\gamma$ pair, time-step and value of the smooth matter fraction $s$, we measured the scatter by taking half the difference of the 16th and 84th percentile of the resulting probability density function (PDF)\footnote{Equivalent to calculating a 1$\sigma$ standard deviation for a Gaussian distribution.}.

In Figure \ref{GD1scattermaps}, we show how the microlensing scatter $\sigma_\mathrm{ML}$ varies with $\kappa$, $\gamma$ and $s$, across a range of times $t$. We find that there is a region of parameter space where the light curves from GLSNe Ia are standardisable as the scatter due to microlensing is comparable to the typical intrinsic dispersion for a SN Ia after standardisation (\citealt{Betoule2014}; \citealt{Macaulay2017}). Therefore, with the correct lensing configuration, it is possible to infer the unlensed magnitude of the source SN Ia. This will lift the mass-sheet degeneracy and allow us to acquire an accurate, precise and independent measurement of the Hubble Constant $H_0$. 

We find that the standardisable region corresponds to lensed images with low $\kappa$ and $\gamma$, with the size of the standardisable region increasing with $s$, i.e. images forming in regions of lower stellar density are less susceptible to microlensing. {Physically, this correpsonds to a lens with an asymmetric image configuration, with at least one image located far outside the Einstein radius of the lens. This outermost image experiences the least amount microlensing due to being far away from the high stellar density region of the lensing galaxy and hence could be used to infer the unlensed magnitude of the background SN Ia.}

{As highlighted in Figure \ref{GD1Scatter}, the microlensing scatter in low $\kappa, \gamma$ regions increases over time, meaning early time measurements of the SN image fluxes are optimal for cosmography. This counter-intuitive result is because there are few caustics in these situations and a small source will typically fall in the smooth region between caustics (see Figure \ref{GERLUMPHExample}). A small number of systems will be highly magnified but these are excluded by our choice to define width as half the 68\% confidence region. As the source expands it is more likely to cross a caustic, creating a larger spread of magnifications at late time. The scatter does not decrease at late times as the source is still too small to average over many caustics. However, in higher $\kappa$, $\gamma$ regions, the scatter decreases with time and increases with $s$. This behaviour is due to the increased density of microlensing caustics. As the source expands it averages over more caustics and the scatter shrinks, but this still does not reach a standardisable level even after 200 days.}

\subsection{How many lensed supernovae are standardisable?} \label{sec:howmany}
To investigate the fraction of lensed supernovae that will be standardisable, we first use a simple lens model to relate lensed image position to the expected microlensing scatter. We assume a {singular isothermal sphere (SIS)} lens model and iPTF16geu redshifts of $z_l = 0.216$ and $z_s = 0.409$ for the lens and source respectively.

For each point on the image plane $\kappa$ and $\gamma$ can be inferred from the macro lens model, however the macro model is sensitive only to the total mass and not the partition between stars and dark matter. The smooth matter fraction is given by the fraction of the surface density not in stars:
\begin{equation}
s = 1 - \frac{\kappa_{*}}{\kappa_\mathrm{tot}}
\label{eq:frac_smth}
\end{equation}
The total mass distribution is modelled using an SIS lens profile:
\begin{equation}
\kappa_\mathrm{tot} = \kappa_\mathrm{SIS} = \frac{R_{\mathrm{Ein}}}{2r},
\label{eq:k_SIS}
\end{equation}
where r denotes a position in the image plane in polar coordinates. For the stellar component of Equation \ref{eq:frac_smth} we assume a de Vaucouleurs profile:
\begin{equation}
\kappa_* = \kappa_\mathrm{deV}(r) = Ae^{-k(r/R_\mathrm{e})^{1/4}},
\label{eq:k_deV}
\end{equation}
where $A$ is a normalisation constant, {$R_\mathrm{e}$} is the effective radius of the lens and $k$=7.669 \citep{Dobler:2006wv}. To calculate the normalisation constant A, we match the dark matter fractions to those found in typical strong lensing ellipticals in the Sloan Lens ACS (SLACS) sample\\ \citep{SLACS_X}: 
\begin{equation}
A = (1 - f_\mathrm{DM})\frac{M_\mathrm{tot}}{M_*},
\end{equation}
where $M_\mathrm{tot}$ and $M_*$ were inferred by integrating equations \ref{eq:k_SIS} and \ref{eq:k_deV} in polar coordinates from 0 to {$R_{e}/2$}. $f_\mathrm{DM}$ is the total projected fraction of dark matter within half the effective radius of the galaxy. Our $f_\mathrm{DM}$ is then matched to the fit derived in \citet{SLACS_X}:
\begin{equation}
f_\mathrm{DM} = a \times \log(\sigma_{R_{e}/2}) + b,
\end{equation}
where $a$ and $b$ are fitting parameters that depend on the initial mass function (IMF) of the lens and {$\sigma_{R_{e/2}}$}\footnote{In units of $100$ km s$^{-1}$.} is the velocity dispersion within half the effective radius of the lens \citep{SLACS_X}. Assuming a Salpeter IMF, $a = 0.80 \pm 0.44$ and $b = −0.05 \pm 0.18$ while for a Chabrier IMF, $a = 0.46 \pm 0.22$ and $b = 0.40 \pm 0.09$ \citep{SLACS_X}. 

{Using our model for $\kappa$, $\gamma$ and $s$ across the image plane and the results of section \ref{sec:Micro_Scatter}, we determine the microlensing scatter as a function of GLSN image plane position}. This is shown in Figure \ref{ImgScatter}. Figure \ref{ImgScatter} shows that standardisable images form beyond the Einstein radius corresponding to an {asymmetric configuration}. More of the image plane is standardisable if the Einstein radius is large or if the IMF is Chabrier rather than Saltpeter. The Chabrier IMF has a lower mass-to-light ratio, so places a larger fraction of the total mass in dark matter whereas the Salpeter IMF places more mass in low mass stars.

\begin{figure}
\includegraphics[width=\columnwidth]{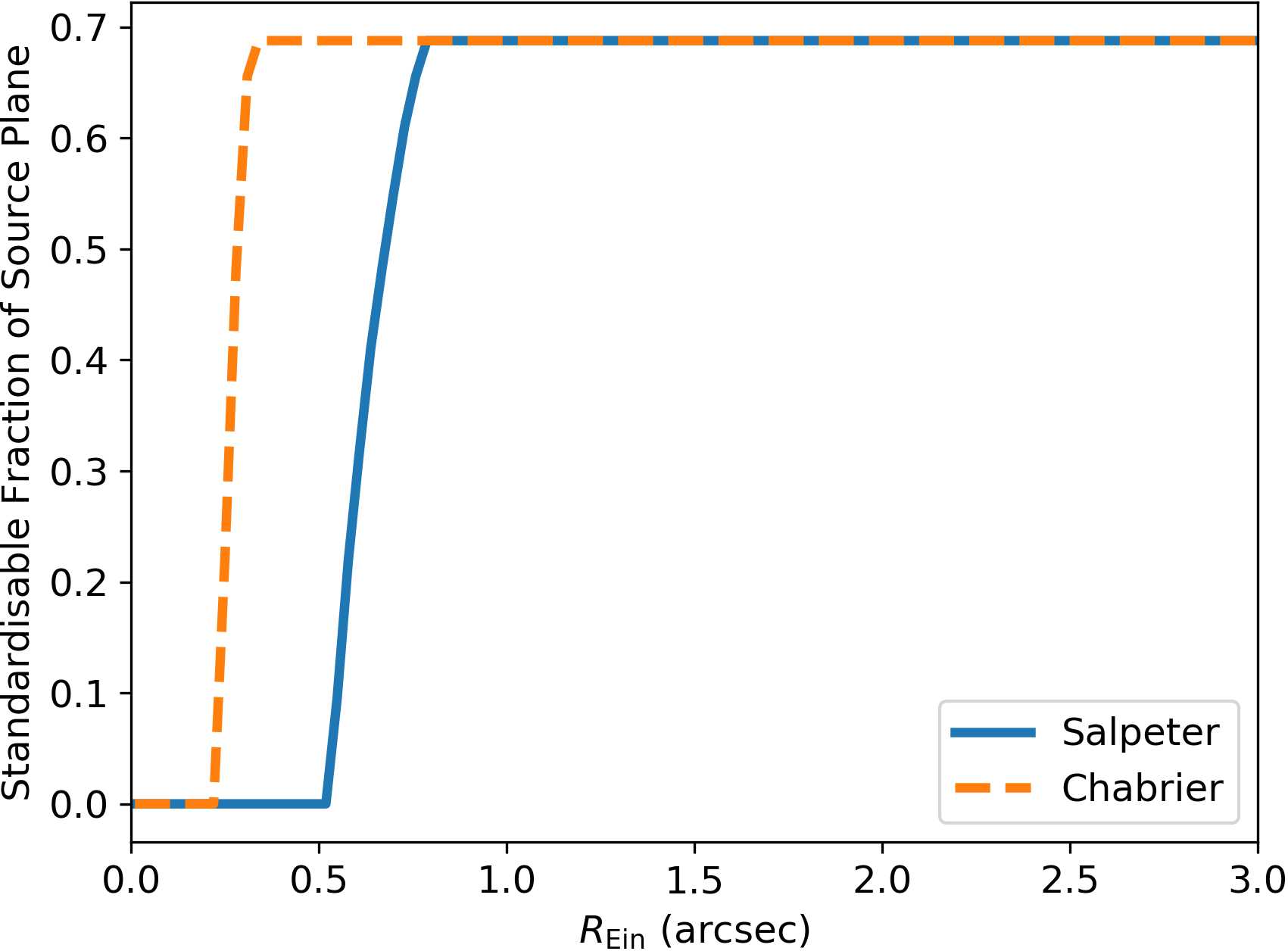}
\caption{The percentage of the source plane that is standardisable as a function of the Einstein radius $R_{\mathrm{Ein}}$ of the lens 30 days after explosion. The result is sensitive to the dark matter fraction in the lens: the solid blue and dashed orange lines correspond to the dark matter fractions derived in \citet{SLACS_X} for a Salpeter and Chabrier Initial Mass Function (IMF) respectively.}
\label{StandFrac}
\end{figure}

{In order to infer how often standardisable images form we must determine the fraction of the source plane that is standardisable. We solve the lens equation for a range of source positions and Einstein radii. For each source position we infer the microlensing scatter for all images formed. Figure \ref{StandFrac} shows the fraction of the source plane that is standardisable as a function of $R_{\mathrm{Ein}}$, for a SN 30 days after explosion, assuming either a Salpeter or Chabrier IMF. For a Salpeter IMF {$\sim$ 70\%} of the source plane is standardisable provided $R_{\mathrm{Ein}} \gtrsim 1$ arcsecond on the sky. Decreasing $R_{\mathrm{Ein}}$ causes the standardisable fraction of the source plane to sharply decline to zero at $R_{\mathrm{Ein}} \sim 0.5$ arcseconds. More of the source plane is standardisable if the IMF is Chabrier: lenses as small as $R_{\mathrm{Ein}} \sim 0.4$ arcseconds can have a source plane which is {70\%} standardisable, but sharply dropping to 0\% at $R_{\mathrm{Ein}} \sim 0.2$ arcseconds. 
\begin{figure*} \centering
\includegraphics[width=\textwidth]{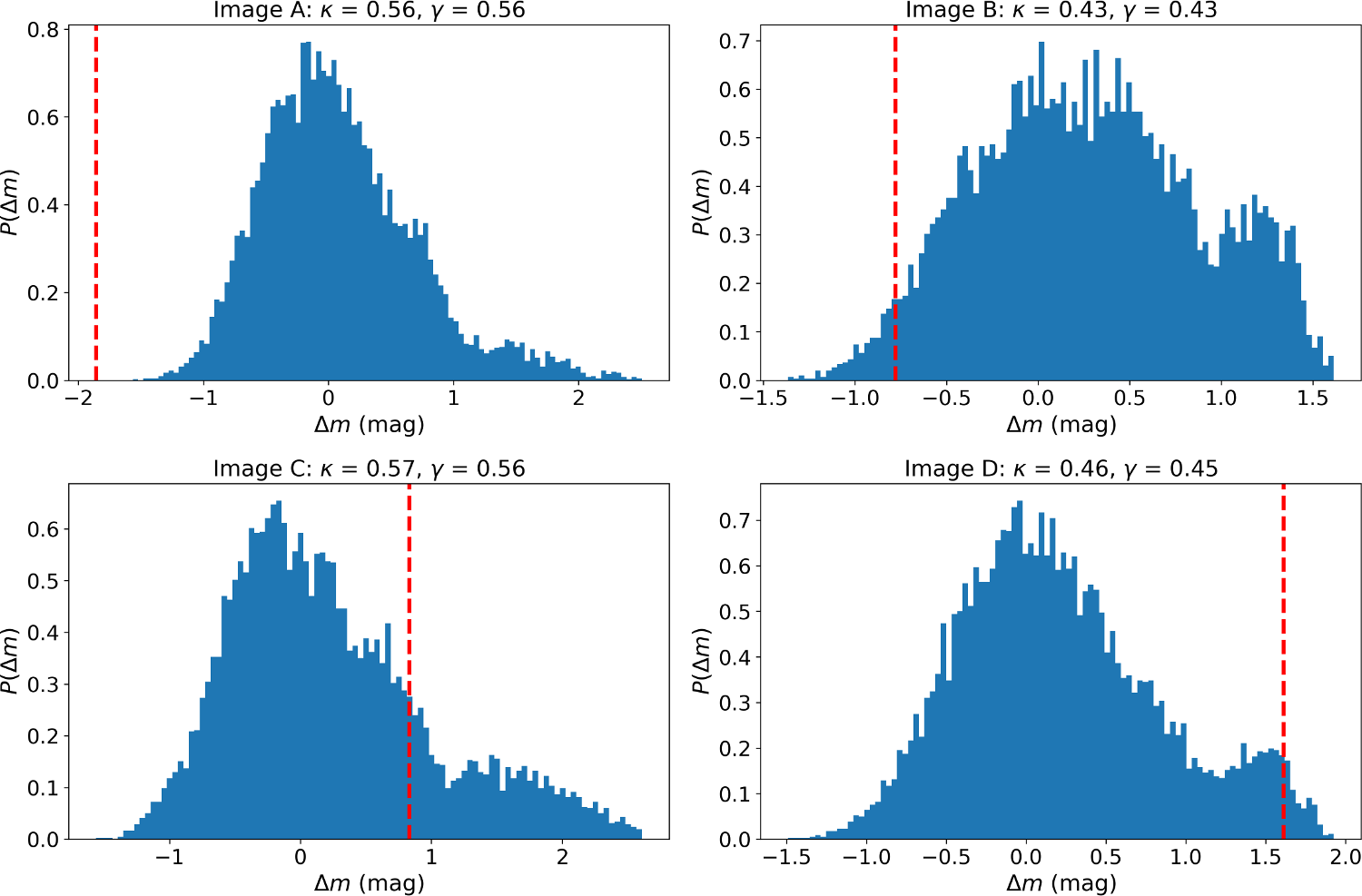}
\caption{Histograms showing the typical magnifications for $10^4$ microlensed supernovae for each image in iPTF16geu, around the time of Hubble Space Telescope (HST) imaging. The magnifications are in units of magnitudes and show the microlensing deviation from the smooth macromodel. The red dashed lines give the corresponding observations from \citet{Goobar:2016uuf}.}
\label{iPTFhistplots}
\end{figure*}
In principle, measuring the scatter for a sample of GLSNe Ia with $R_{\mathrm{Ein}} \sim$ 0.5 arcseconds will allow us to  discriminate between IMFs in the lensing galaxy. If the \citet{SLACS_X} Salpeter fit is correct then no lensed SNe with $R_{\mathrm{Ein}}<0.5$ arcseconds should have a scatter of less than 0.15 mags whilst most lensed SNe will if the Chabrier fit is correct.

The above toy model neglects magnification bias. Whilst {asymmetric configurations} dominate the source plane, they are less highly magnified and therefore harder to detect than more {symmetric configurations}. To illustrate this we take the mock GLSNe catalogue of \citet{GoldsteinTimeDel2017} and assess the standardisable fraction of systems where the brightest SN image reaches a peak apparent i-band magnitude of 22.15. This choice roughly approximates the LSST discovery threshold for GLSNe. We predict that that 22\% of the $\sim$ 930 GLSNe Ia {to be} discovered by LSST will be standardisable, {of which approximately 1 in 5 will be quads.} The median time delay for a standardisable LSST GLSN Ia is 44 days, compared to 18 days for all LSST GLSN Ia. The median Einstein radius for a standardisable LSST GLSN Ia is 0.9 arcseconds, compared to  0.7 arcseconds for all LSST GLSN Ia. {The catalogue spans a range of Einstein radii between 0.06 arcsec $\leq R_\mathrm{Ein} \leq$ 2.54 arcsec}.

\subsection{Lifting the mass-sheet degeneracy with LSST GLSNe Ia: predictions for $H_0$}

The fundamental gain of a GLSNe Ia over a standard time delay lens is the ability to test for the presence of systematic uncertainties in the lens model. Since lens models have typical errors of a few percent \citep{H0LiCOW42017}, a 0.15 mag uncertainty on the flux of a single lensed image will not provide statistically relevant improvement on $H_0$. Averaging over many lensed supernovae will be required to constrain $H_0$ with interesting accuracy. 

To investigate the expected constraints on $H_0$ from the final LSST GLSNe Ia sample, we draw 650 GLSNe Ia from our mock LSST catalogue. This is the number of GLSNe Ia forecast to be discovered by LSST early enough to measure reliable time delays \citep{GoldsteinTimeDel2017}. Taking a typical 7\% error per system \citep{H0LiCOWV2017} and scaling by root N, gives {$\sigma_{H_0} = 0.3\%$}, however this neglects residual systematics from the mass-sheet degeneracy.

Given the individual {magnification probability} $P(\mu)$ for each image of a GLSN, the expected PDF for the unlensed magnitude of the supernova is given by:
\begin{equation}
P(M_{\mathrm{SN}})= \prod_{\forall i} P(\mu_i).
\end{equation}
Adding this in quadrature to an intrinsic {SN Ia} scatter of 0.1 magnitudes \citep{Betoule2014}, gives the expected uncertainty on the macromodel magnifications for the lens. 

Constraining the true macromodel magnifications gives constraints on the mass-sheet degeneracy parameter $\lambda$, since 
\begin{equation}
\mu_\mathrm{True} = \mu_\mathrm{Model}/\lambda^2,
\label{eq:degeneracymu}
\end{equation}
where $\mu_\mathrm{True}$ is the true magnification and $\mu_\mathrm{Model}$ is the macromodel magnification assuming $\lambda=0$. This implies that a system with a microlensing scatter of 0.15 magnitudes gives a constraint on $\lambda$ with 17\% precision that is insensitive to the the mass-sheet transformation.

The time delays - and hence $1/H_0$ - are proportional to $\lambda$. The product of $P(\lambda)$ over all the systems gives the level at which the mass-sheet degeneracy can be broken for the final constraint on $H_0$. This product has a width of {$\sigma_{\lambda} = 0.5\%$}. Combined, the 650 LSST GLSNe Ia will therefore be able to detect systematics in $H_0$ due to the mass-sheet transformation at the 0.5\% level. If we restrict the sample to the 140 GLSNe with a microlensing scatter of less than 0.15 mags, the constraints on $H_0$ degrade to {$\sigma_{H_0} = 0.6\%$}. {If only the 44 standardisable quad image systems are used, the constraints $H_0$ degrade to {$\sigma_{H_0} = 1.1\%$}}.

\subsection{iPTF16geu}
\begin{table}
\caption{Table of parameters used for iPTF16geu simulations. $\kappa$ and $\gamma$ were taken from \protect\cite{More:2016sys}. Values of the smooth matter fraction $s$ were inferred from our lens model of iPTF16geu.}
\label{table:iptf}
\begin{tabular}{c|ccc}
Image & $\kappa$ & $\gamma$ & $s$\\
\hline
A & 0.56 & 0.56 & 0.22 \\
B & 0.43 & 0.43 & 0.23 \\
C & 0.57 & 0.56 & 0.27\\
D & 0.46 & 0.45 & 0.23
\end{tabular}
\end{table}
The recent observations of iPTF16geu, give us a first opportunity to test the analysis methods developed in this paper. Since the images have high magnification and form in regions of high stellar density, we should not expect this system to be standardisable.
The values of $\kappa$ and $\gamma$ for each of the iPTF16geu images have been estimated from the macro lens model published in \citet{More:2016sys}. We use the same prescription as in Section \ref{sec:howmany} to infer the likely smooth matter fractions at the image locations (Table \ref{table:iptf}).

In Figure \ref{iPTFhistplots} we show the PDF of the change in magnitude due to microlensing for each image of iPTF16geu. These PDFS are generated assuming a time of 60 days after explosion approximately corresponding to the HST data analysed in \citet{More:2016sys}. \citet{More:2016sys} noted that there is a significant discrepancy between the observed fluxes in iPTF16geu and those predicted by their lens macromodel assuming iPTF16geu is a typical SN Ia, with image A being almost 2 magnitudes brighter than the macromodel prediction for a SN Ia. Figure \ref{iPTFhistplots} shows that this discrepancy cannot be due to microlensing alone. A similar analysis by \citet{Yahalomi2017} reaches the same conclusion using a point source, however the tension increases for a finite sized source. We find that the discrepancies between the observed and macromodel predicted fluxes of the other three images are consistent with microlensing.

If iPTF16geu has a standard type Ia luminosity, then at least some of the the disagreement in the observed and predicted fluxes must be due to deficiencies in the macromodel. The presence of a dark substructure or a stellar disk close to image A may explain this flux anomaly \citep{vegetti2010,Hsueh2017}. 

iPTF has an r-band discovery limit of 21st {magnitude} \citep{Goobar:2016uuf}; without the extreme magnification of image A, iPTF16geu {still would have been identified as a transient by iPTF, but only marginally. The transient was only added to the spectroscopic follow-up queue when the system reached an r band magnitude of 19.3 \citet{Goobar:2016uuf}. iPTF16geu would likely had not been confirmed as a GLSNe Ia without the extreme magnification of image A. The demagnification of Image D is another atypical feature of iPTF16geu. Whilst microlensing can plausibly explain the observed brightness of D, the presence of dust may also contribute to the dimming. Therefore,} the micro- and macro-lensing of iPTF16geu are therefore unlikely to be representative of a future population of lensed SNe. However this result does demonstrate that breaking the mass-sheet degeneracy with future lensed SNe Ia will also require a detailed reconstruction of disks and dark matter substructures in the lenses. 

\section{Conclusions} \label{sec:conclusion}

We have evaluated the effect of microlensing on GLSNe Ia for various image configurations, corresponding to values of the convergence $\kappa$, the shear $\gamma$ and the smooth matter fraction $s$, across multiple time intervals. We have found that there are regions of parameter space where the effect of microlensing is suppressed enough for the GLSN Ia to be standardisable. Specifically, regions of low $\kappa$, $\gamma$ and high $s$ are subject to microlensing scatter of $\sigma_\mathrm{ML} \lesssim 0.15$, particularly at early times (see Figure \ref{GD1scattermaps}). {Physically this corresponds to asymmetric configurations with at least one image located far outside the Einstein radius}, which will experience the least amount of microlensing.}

Combining our microlensing models with the GLSNe Ia catalogue from \citet{GoldsteinTimeDel2017}, we predict that $\sim$ 22\% of the $\sim$ 930 GLSNe Ia {to be} discovered by LSST will be standardisable. From the sample of 650 GLSNe Ia, of which accurate time delays can be measured, the mass-sheet degeneracy can be broken at the 0.5\% level. The LSST GLSNe Ia sample will thus be robust against systematics in $H_0$ at the 0.5\% level. The assumed fraction of standardisable systems with accurate time-delays may be somewhat pessimistic, since we found that standardisable GLSNe have larger Einstein radii (median 0.9 arcseconds) and time delays (median 44 days), than the general population. 

Our result assumes a SN Ia light profile that expands at a constant velocity of $10^4$ km s$^{-1}$. Whilst simple, more complicated models can be extracted from our results by rescaling the time axis. {We have not considered sources with clumpy profiles, however, since the standardisable region of Figure \ref{GD1scattermaps} varies only weakly with time, our results should not be heavily influenced by the choice of source model. If the supernova profile contains any small, bright, fast moving clumps then additional scatter may be introduced. However, microlensing of such clumps would introduce rapid temporal variation in the light curve which should be easy to detect.}

Whilst this paper does not focus on the IMF, we found a sharp sensitivity to the IMF for lenses with Einstein radii between 0.2 and 0.5 arcseconds, assuming a lens and source with the same redshifts as in iPTF16geu. Measuring the scatter in a sample of such GLSNe Ia should trivially discriminate between the Salpeter and Chabrier fits of \citet{SLACS_X}.

We also applied our microlensing analysis to the GLSN Ia iPTF16geu and compared our results against the \citet{More:2016sys} analysis, who found a strong discrepancy between the observations and their lens model, attributing the discrepancy to microlensing. Our analysis suggests that the discrepancy cannot be due to microlensing {primarily} (see Figure \ref{iPTFhistplots}) and signals potential deficiencies in the {use of} simple lens macromodels, as suggested by \citet{More:2016sys}. 

This work shows that it is possible to infer the intrinsic luminosity for a significant sample of {$\sim$ 200 LSST} GLSNe Ia, suppressing the mass-sheet degeneracy of the lens model. This will allow for accurate and precise measurements of $H_0$ with significantly reduced systematics through time-delay cosmography, thus enabling a stringent test of the $\Lambda$CDM model of cosmology.

\section*{Acknowledgements}
We thank Bob Nichol, Daniel Whalen, Daniel Thomas, Paul Schechter, Peter Nugent {and the referee} for constructive discussions of this paper.

We are grateful to the Royal Society for an International Exchange Grant (IE170307) which has supported our collaboration on this work.

MF is supported by the University of Portsmouth, through a University Studentship. 
TEC is supported by a Dennis Sciama Fellowship from the University of Portsmouth. 
GV is supported through an NWO-VICI grant (project number 639.043.308). 
DAG acknowledges support from the DOE under grant DE-AC02-05CH11231, Analytical Modelling for Extreme-Scale Computing Environments. 
DB is supported by STFC consolidated grant ST/N000668/1. This work was performed on the gSTAR national facility at Swinburne University of Technology. gSTAR is funded by Swinburne and the Australian Government's Education Investment Fund.
%%%%%%%%%%%%%%%%%%%%%%%%%%%%%%%%%%%%%%%%%%%%%%%%%%



%\nocite{*}
\bibliographystyle{mnras}
\bibliography{ref} % if your bibtex file is called example.bib

%%%%%%%%%%%%%%%%%%%%%%%%%%%%%%%%%%%%%%%%%%%%%%%%%%

%%%%%%%%%%%%%%%%% APPENDICES %%%%%%%%%%%%%%%%%%%%%

%\appendix



%%%%%%%%%%%%%%%%%%%%%%%%%%%%%%%%%%%%%%%%%%%%%%%%%%

% Don't change these lines
\bsp	% typesetting comment
\label{lastpage}
\end{document}